# Stacked Multilevel Fresnel Zone Plates for Hard X-Rays

L. Haroutunyan[a], K. Trouni[b], A. Kuyumchyan[c,d]

[a] Yerevan State University, Alex Manoogian St. 1, Yerevan 375025, Armenia. E-mail: levon.har@gmail.com
[b] Candle Synchrotron Research Institute Foundation, Acharyan St. 31, Yerevan 375040, Armenia
[c] Institute of Microelectronics Technology, RAS, Academician Ossipyan St. 6, Chernogolovka 142432, Russia
[d] American NanoScience and Advanced Medical Equipment, Inc., CA 91204, USA

The new type of multilevel Fresnel zone plate, consisting of stacked layers with bi-level zone profile has been investigated. The conditions of layers acting as single multilevel Fresnel zone plate have been discussed by numerical simulation. The criteria for layers alignment have been presented. Considered approach of zone plate fabrication allows achieving high level focusing efficiency from one side and decreasing zone profile etching deep from the other.

## 1. Introduction

Fresnel zone plates (FZP) [1] have been widely used as focusing elements in hard X-ray microscopy for the last two decades. The theoretical limit of focusing efficiency of phase-shift FZP is $\eta = 4/\pi^2 \cong 40.5\%$. More than a decade ago, in order to increase the focusing efficiency the so-called multilevel FZPs were created [2], which was a better approximation for ideal focusing structure – kinoform. Each couple of zones of bi-level FZP has been replaced by $M$ zones with phase-shift growing by $2\pi/M$ at each subsequent zone. For non absorbing system only the $j = Mk + 1$ $(k = 0, \pm 1, \pm 2, \dots)$ orders of diffraction are active. The focusing efficiency of the first order diffracted waves is $\eta_1 = \sin^2(\pi/M)/(\pi/M)^2$ and decreases by factor $1/j^2$ for higher orders. Even for three level FZP ($M = 3$) the first order focusing efficiency increases up to 68.4% from above mentioned 40.5% for bi-level FZP. Despite of obvious advantage, the practical application of such devices is restricted due to difficulties in their fabrication.

A new type of multilevel FZP consisting with series of closely spaced layers with common optical axes – the so called stacked multilevel FZP will be considered in this paper. Each layer of proposed device has bi-level profile with $2\pi/M$ phase shifting as shown in fig.1a. There $r_n = \sqrt{2\lambda Fn/M}$ is the nth zone external radius of multilevel FZP, $\lambda$ - radiation wave length, $F$ – FZP's first order focus distance, $M$ - levels number, $l$ - interlayer distance.

For radiation falling along optical axes, the wave diffraction in interlayer jump can be neglected if $l$ is sufficiently small. Then we will have a profile of multilayer FZP as a result of simple superposition of layers profiles (fig.1b). Besides of high focusing efficiency characteristic for multilevel FZP, another advantage of the considered scheme is the low height of zones profile compared with the bi-level zone plates. This is important from the experimental point of view, as it allows decreasing the outermost zone width and consequently increasing the focus sharpness.

A bi-level stacked FZP for high energy X-rays focusing was already considered by numerical simulation [3] and experimentally [4-6]. The aim was to decrease necessary etching depth of zones, which is especially important for higher photon energy. The acceptable values of interlayer distance and radial displacement of two layers for which they still act as one zone plate with doubled phase shift was found. The criterion of layers radial alignment has been developed as well. It is natural that our consideration developed mainly on the analogue way but for multilevel zone plates.

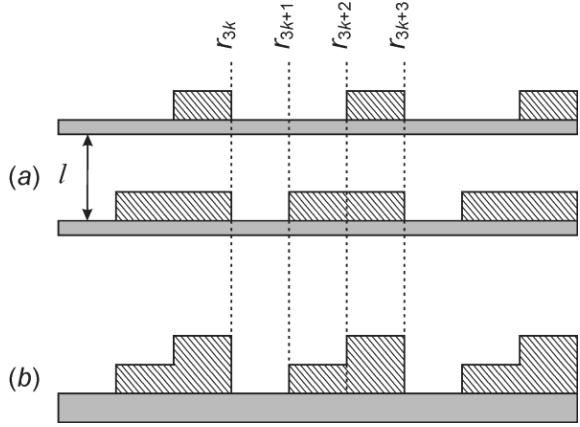
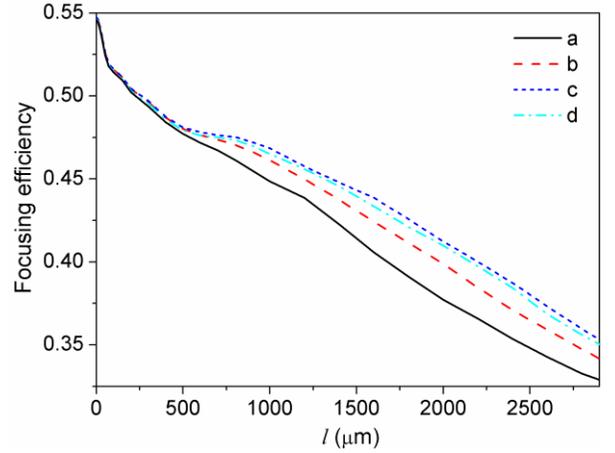

Fig.1. Fragment of stacked multilevel FZP for $M = 3$ case.
a) Layers and their relative position.
b) Projection of layers zones profile.

Fig.2. Focusing peak efficiency dependence from interlayer distance $l$ for incident plane wave (a), spherical waves with object distances $5F$ (b), $2F$ (c), and $1.5F$ (d).

## 2. Interlayer distance

In purpose to define acceptable value of interlayer distance $l$, the numerical simulation of X-ray diffraction in stacked multilayer FZP was carried out for different $l$ and analyzed the dependence of focusing properties from $l$. Plane wave parallel to optical axes, as well as spherical wave from point source located on optical axes at different object distances ($d = 5F; 2F; 1.5F$) as initial radiation was discussed. The following values of the main parameters involved in simulation were used: $\lambda = 0.1$ nm, $F = 1$ m, number of zone rings $N = 210$, two silicon layers case was considered. In this case zone plate radius $r_N = 118.3$ μm, outermost zone width $\Delta r_N \equiv r_N - r_{N-1} = 0.282$ μm, layers zone profile height $h = 10.53$ μm. The calculated dependence of focusing efficiency on interlayer distance $l$ was shown in fig.2[1]. First of all the quite weak dependence of efficiency from incident radiation wave front curvature must be mentioned. As expected, the efficiency reaches its highest value for $l = 0$ and decreases by increasing $l$. For $l$ up to several hundred μm the drop of efficiency does not extend 5-10%. For example, in case of incident plane wave, increase of $l$ from 0 up to 500 μm, decreases the efficiency from 54.6% to 47.7%, which is still higher from that of bi-level FZP even for non absorbing layers (in future calculations we consider $l = 500$ μm case, as we hope it is quite realizable experimentally). Considered dependence is in good agreement with simple analytical estimations. Diffraction in interlayer region does not play a significant role (near field approximation) in case of diffraction widening of wave passed from the narrowest zone is smaller compared to the zone width: $(\lambda/\Delta r_N)l < \Delta r_N$ or

$$l < \frac{(\Delta r_N)^2}{\lambda} = \frac{F}{6N}. \qquad (1)$$

In considered case it follows that $l < 800$ μm. Condition (1) matches qualitatively with one in [3].

---

[1] Here in numerical calculations contributions of focus peak satellites didn't included in efficiency, in contrast to above mentioned analytical estimation. This and partly limited accuracy of numerical calculations leads to some difference between numerically simulated and analytically estimated efficiencies.

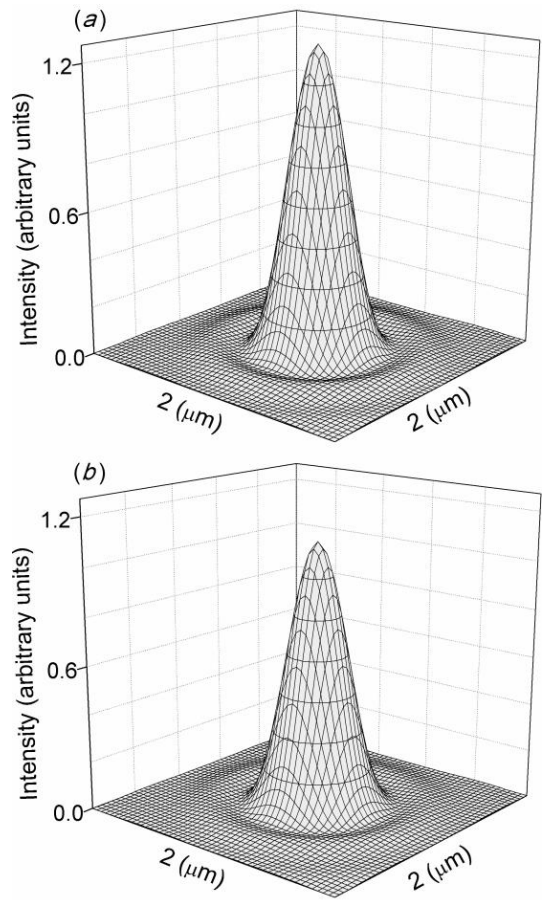

Fig. 3. 2D intensity map of focus peak for interlayer distances $l = 0$ μm (a) and $l = 500$ μm (b).

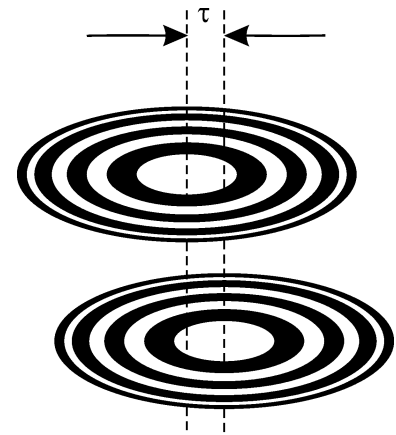

Fig.4. Schematic drawing of radial displaced layers.

The peak intensity spatial distributions for interlayer distances $l = 0$ μm and $l = 500$ μm are presented in fig.3[2]. As can be seen from fig.3 the presence of interlayer distance decreases peak height without considerable changes in width and form.

## 3. Layers radial displacement

Up to this point it was assumed that optical axes of layers coincide exactly, which, of course is inaccessible from experimental point of view. In this section we will consider the influence of layers radial displacement $\tau$ on focusing properties (fig.4) and search for criteria, which can be used for layers arrangement. The numerically simulated intensity distributions in neighborhood of focal point for different values of $\tau$ are presented in fig.5. As can be seen from drawing, focus peak splits in double peaks with appearing of secondary or maybe multiple maximums by increasing $\tau$. The distance between sub-peaks approximately equal to $\tau$ and the total power of all peaks is significantly less than that of single peak. The dependence of focusing efficiency[3] from $\tau$ is shown in fig.6. According to that $\tau_m \approx 0.1 \div 0.15$ μm can be considered as maximum radial displacement in which focusing properties aren't significantly disturbed. This matches to appropriate estimation for bi-level stacked FZP according to which $\tau_m \approx \Delta r_N/2$ (note the difference of zone widths in bi- and three-level FZP's).

The same experimental criterion, as that in case of bi-level stacked FZP, can be used for layers radial alignment. It is based on the fringes formation behind stacked FZP due to radial displacement. Calculated intensity distribution at image distance $f = 2F$ for different radial displacement $\tau$ is shown in fig.7. As can be seen, the radial displacement arise an interference fringes perpendicular to horizontal directed displacement. The fringes period decreases by increasing $\tau$.

---

[2] Here and later only plane wave parallel to optical axes was considered as incident radiation, unless is not mentioned otherwise.
[3] Contribution of all peaks is taken into account in efficiency.

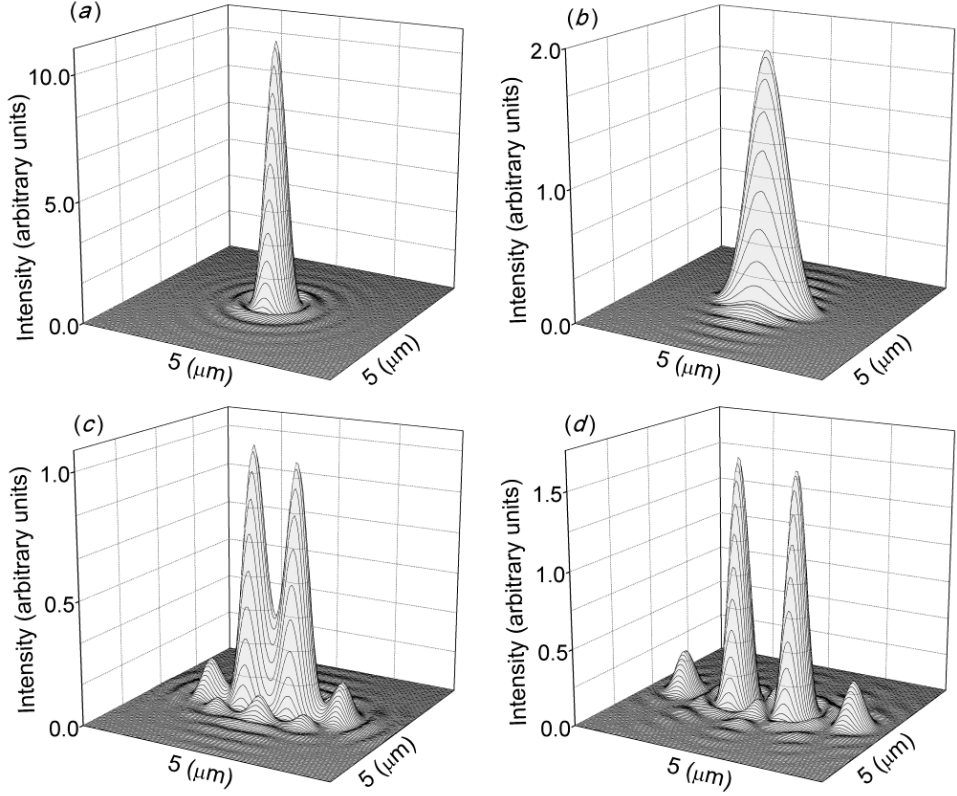

Fig.5. Intensity distribution near focus for 0 (a), 0.5 μm (b), 0.8 μm (c), 1.2 μm (d) radial displacement $\tau$.

As an alternative criterion for layers radial alignment the efficiency of 0-order diffracted radiation can be served. As mentioned above 0-order diffraction isn't active for pure phase multilevel FZP. Of course absorption in zone structure, as well as diffraction in interlayer gap of stacked multilayer FZP leads to the appearance of 0-order diffraction, but its efficiency is still low in considered case. Meanwhile layers radial displacement being in order of $\Delta r_N$ increases 0-order diffraction efficiency substantially, this can be served as criterion for radial alignment. The power of 0-order diffraction may be estimated by registering the radiation passing through the circular pinhole with radius $r_N$ placed at the distance $f \geq F$ from zone plate, meantime suppressing strong first-order diffraction by a tiny stopper on the focus point. For incident plane wave 0-order diffracted radiation propagates parallel and passes through the pinhole almost entirely. On the other hand only small parts of other order diffracted radiation pass through the pinhole as they are either always divergent (negative orders) or focusing in $f < F$ and diverging after (positive orders).

The calculated dependence of registered efficiency from $\tau$ for different FZP–pinhole distances $f$ and corresponding optical scheme of measurement are shown in fig.8. As one can see from graphics, the sharpness of power drop near $\tau = 0$ increases by increasing $f$.

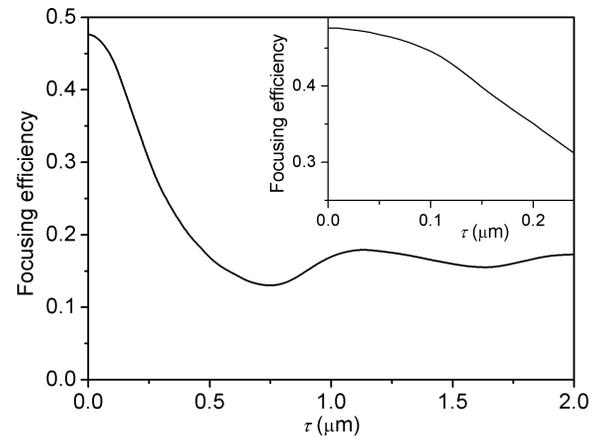

Fig.6. Focusing efficiency dependence of layers radial displacement $\tau$.

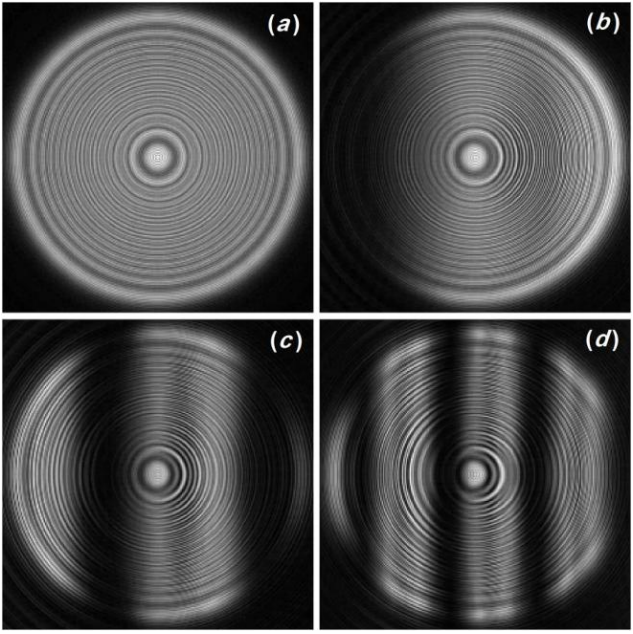

Fig. 7. Calculated intensity distribution at image distance $f = 2F$ for different radial displacements (a. $\tau = 0$; b. $\tau = 0.3$ μm; c. $\tau = 0.8$ μm; d. $\tau = 1.5$ μm).

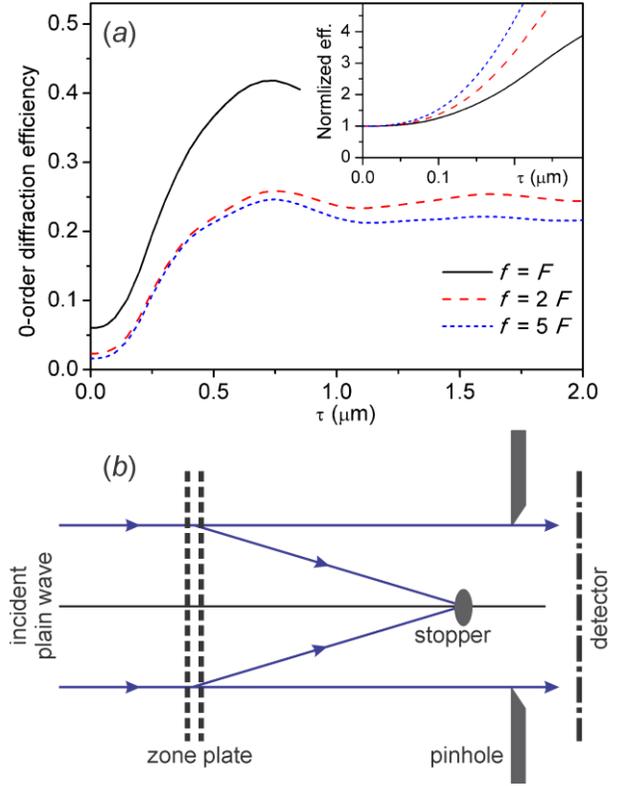

Fig.8. Approached 0-order diffraction efficiency dependence from $\tau$ for different FZP–pinhole distances (a) and registration optical scheme (b). Efficiencies normalized by their values at $\tau = 0$ are plotted in right-higher corner of the graph.

Although calculations confirm the usefulness of the scheme described, a notice must be made. Due to the limits of fabrication technique, zone profile often doesn't have the strictly required form. As a result, the mentioned sharp drop of the 0-order diffraction efficiency around $\tau = 0$, can be smoothened. It is reasonable to assume, that main imperfections of the zone profile corresponds to zones with smaller width and higher radius consequently. Hence, depending on imperfections character, their influence can be reduced by decreasing the pinhole radius or by inserting an additional pinhole with smaller radius straight before or after FZP.

## 4. Sensitivity to angular deviations

Perhaps the main disadvantage of considered device compared with ordinary multilevel FZPs and multilevel-type multilayer FZPs [7-8] is its high sensitivity to angular deviation of incident radiation from optical axes. Suppose radiation point source S shifted from optical axes by $\rho$ (fig. 9). This shifting is equivalent to zone plate layers rotation in $\varphi = \rho/d$ angle and layers radial displacement $\tau = \varphi l$, where $d$ – object distance. Being described by the angular sensitivity of ordinary FZP the first factor is negligibly small in the considered case. Taking into account that $\tau$ should not exceed $\tau_m \approx \Delta r_N/2$, second factor leads to the condition $\rho < \rho_o$. Here $\rho_0 = d\Delta r_N/(2l)$ is the radius of the circle in object plane with acceptable quality imaging. According to (1) and simple relation $\Delta r_N \cong r_N/(2N)$ $\rho_0$ can be estimated from below $\rho > 1.5 r_N \, d/F > 1.5 r_N$. Hence relatively high angular sensitivity of stacked multilevel FZP still allows imaging objects larger than zone plate itself. As an example the imaging of grid structure with narrow transparent slits has been simulated in fig. 10.

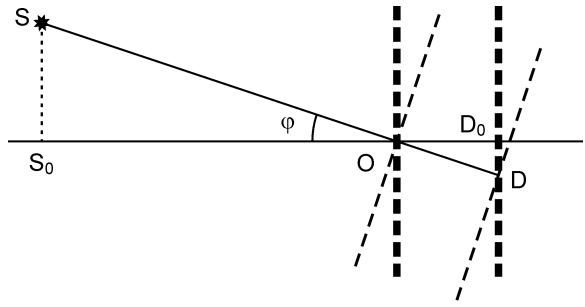

Fig.9. Schematic diagram in the case of incident radiation point source shifted from optical axes. S: point source, $d = S_0O$, $\rho = SS_0$, $l = OD_0$, $\tau = DD_0$.

## Conclusions

A new type of multilevel Fresnel zone plate named as stacked multilevel FZP, consisting of stacked layers with bi-level zone profiles, was considered by means of numerical simulation. The interlayer gap width and radial displacement of layers at which focusing properties of system are close to that of multilevel FZP, which is the result of simple superposition of layers profiles, was found.

As in case of stacked bi-level FZP, interference fringes formed behind optical system can be used for layers radial alignment. An alternative criterion also was considered based on 0-order diffraction power.

The angular sensitivity of device were observed and shown, that it still allows good quality imaging of objects larger than the FZP diameter. As an example the imaging of grid structure with size equal to FZP diameter has been simulated.

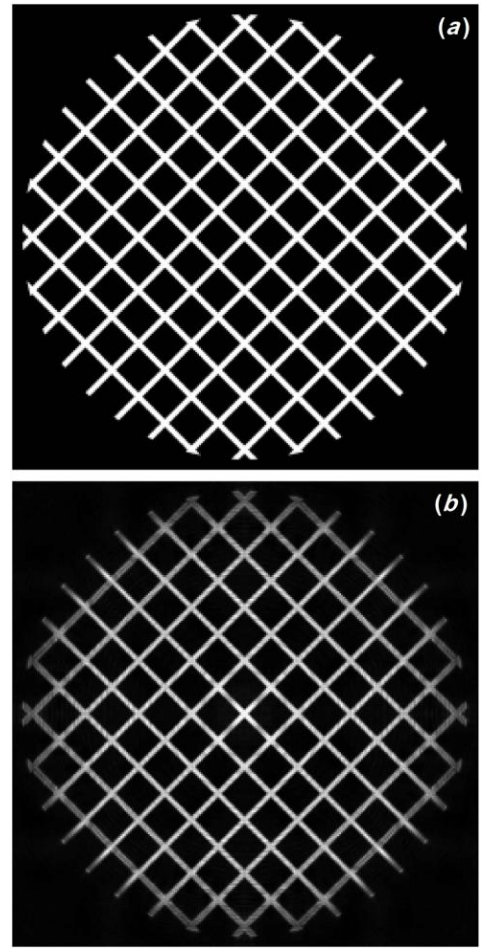

Fig.10. Grid structure imaging with stacked multilevel FZP (a - object, b – image). Object distance: $d = 2F$, grid period – 20 μm, slits thickness – 3 μm, object diameter equals to zone plate diameter.